# Context Awareness in Next Generation of Mobile Core Networks

Clarissa Cassales Marquezan, Huawei Technologies
Kashif Mahmood, Telenor
Dirk Trossen, Interdigital
Anastasios Zafeiropoulos, Ubitech
Renan Krishna, Interdigital
Xiaofeng Huang, Orange
Xueli An, Huawei Technologies
Daniel Corujo, IT Aveiro,
Filipe Leitão, NEC Laboratories
Maria Lema Rosas, Toktam Mahmoodi, King's College London
Hans Einsiedler, Deutsche Telecom

**Abstract**

Context awareness is an important enabler for next generation of Mobile Core Networks (MCN). However there exist a number of challenges in this regard. For example how to develop a framework which 1) is able to generate *context* richer than what is available today; 2) allows reusability of context across the network; 3) provides a mechanism for exposing context to third parties; and 4) can bring together "big data" for mobile core network optimization. In this work, we introduce a context awareness framework addressing the aforementioned challenges but also taking into account the 3GPP standardization activities related to context awareness in MCN. Within this framework we propose Context Generation and Handling Function (CGHF) which generates rich context by processing information from various sources and then handles its distribution through an efficient publish subscribe mechanism. In addition we provide examples where context can be used to optimize control plane decision making. While the focus of this work is on the use of context for MCN, we still believe such context can be also used by applications (at the edge as well as in data centers) and third party services to improve their operations and providing new unforeseen services.

## 1 INTRODUCTION

5G has been described as the nervous system of society, providing the network infrastructure to many verticals with widely varying requirements and constraints, both technologically and economically. In order to address these diverse requirements, system flexibility is one of the key principles [1] [2]. The current introduction of software technology into communication networks, often labelled as network softwarization, is recognized as enabling the flexibility we need through easier upgrade of software components compared to the expensive and inflexible hardware based solutions. Network softwarization along with other enabling concepts such as architectural modularization and network slicing [3] provide the system flexibility envisaged for 5G, yet, these concepts pose a significant challenge of system complexity due to the new increased level of dynamicity. This challenge can only be tamed by converting the communication infrastructure of 5G essentially into a knowledge plane as envisioned in earlier works [4]. However the challenge lies in exploiting the already existent *knowledge* albeit locked in today's networks as well as generating new knowledge. This paper is a step in that direction by proposing a context awareness framework for next generation *Mobile Core Network* (MCN) where the term *context* refers to the knowledge describing the situation of an *entity* [11] that can be a *device, user, network, environment* or *an application*. We next present the challenges towards the realization of such a context awareness framework. Before that it needs to be highlighted here that we limit the application of the proposed framework to the core network in this paper but the framework is generic and can be applied to any part of the 5G network.

First and foremost the context used in today's Long Term Evolution (LTE) networks [5] such as Packet Data Network (PDN), Mobility Management (MM), and Evolved Packet System (EPS) bearer contexts are not rich enough. For instance, the PDN context includes the identity of a PDN gateway and Access Point Name (APN) indicating the gateway and APN being used, while no information regarding the PDN subscription usage and status is provided. This is one of the many examples where the context in todays' network is not so rich, but rather a simple set of what we call raw information, in this paper. One of the reasons hindering the generation of such rich context is that the information is locked in isolation within today's MCNs and its specific Network Functions (NFs) [6] or in application level [7] . This leads to the second challenge of reusability of information and context as the existing frameworks do not enable a cross-NF management of context and has limited exposure to applications. Thirdly, while very recently context exposure has been identified as a key issue in 3GPP [8], yet it is not clear how it can be expose. Therefore, there are no established ways to provide clear business interface. Finally while big data has been used mainly at the application level, e.g., using mobility data of users for predicting the spread of Dengue fever [9]. The use of big data in MCN operation of Control Plane (CP) functions is less explored but presents a lot of potential. This can be supported by the fact that very recently an analytics based policy framework solution is proposed in 3GPP [8] but again that is mainly focuses solely on user data connectivity management and relying on usage of context for a specific NF (i.e., policy framework) which is against the principle of context reusability.

In this paper, we shed light on how to address the aforementioned challenges in an architectural manner through three key contributions. Firstly, we outline the design for a Context Awareness Framework (CAF) that manages as well as handles context knowledge across the different entities of the communication fabric, thereby generating rich and reusable context. It needs to be highlighted that the generated context is a result of inference over ontologies, which moves away from today's narrow implementations of fact based adaptions, e.g., reducing transmission rate at a gateway upon receiving the monitored information about congestion. Secondly, we outline specific examples where rich context, not only raw information from the components themselves, allows for optimization of MCN operations. The examples include context for reevaluation of policy functions, data plane anchor point reselection, service point reselection, and multi-access optimization. Thirdly, we embed our context approach into the wider 5G system by outlining the exposure to third parties and developers via a Northbound Interface (NBI) in order to harvest the innovation potential that lies beyond the optimizations of NFs, and supporting vertical services, applications, and third party services also to optimize their operations. In addition, throughout our paper we also provide an overview of current 3GPP, a key Standardization Organization (SDO) for MCN, limitations and tendencies to support context awareness in 5G MCN.

## 2 NEXT GENERATION NETWORK ARCHITECTURE

Network softwarization trend, propelled mainly by the two key concepts of Network Function Virtualization (NFV) and Software Defined Networking (SDN), intend to transform 5G network [2]. Figure 1 shows a comparison of (a) current LTE architecture and (b) one realization of 5G softwarized network. Figure 1 (b) aims to highlight the differences from Figure 1 (a) by depicting the decoupling of the CP and Data Plane (DP), the SDN principle, decoupling of hardware from software, the NFV principle, and the dynamicity of the next generation networks in which the NFs can be distributed in different parts of the network cloud (public or private) as edge computing or fog computing nodes. This increased level of dynamicity and heterogeneity of entities results in a complex scenario for 5G MCN, posing significant challenges in terms of defining a logical architecture and knowledge acquisition, handling, and utilization, as shown in the studies by SDOs [8][10].

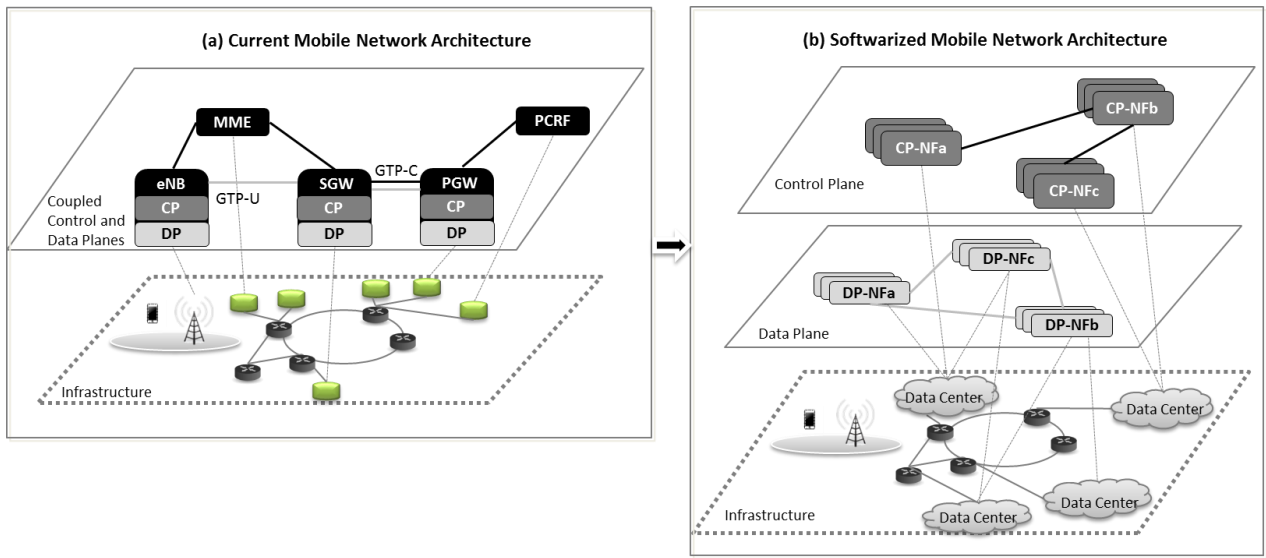

*Figure 1 – A comparison of 4G and 5G mobile architecture. (a) presents the traditional coupled control and data plane were the network functions (NFs) run on purpose built hardware and are mostly static. (b) depicts a dynamic softwarized world leveraging SDN, NFV in which the NFs can be instantiated in different parts of the network.*

In this paper, we take inspiration from the modularized 5G MCN architecture defined by An et al [3]. A simplified version of this architecture is sketched in Figure 2, consisting of both Access Network (AN) and Core Network (CN). For the CN the authors specify dedicated CP NFs such as Connectivity Management (CM), Mobility Management (MM), Flow Management (FM), Security and Authentication Management (SAM), as well as Access Function (AF). It is the Context Generation and Handling Function (CGHF) which is the focus of this paper and the main building block of the CAF. The CAF proposed in this paper leverages the modularization concepts from this architecture, so that reusable context can be generated and consumed by the different NFs aiming at optimizing their decision making, which in turn results in an efficient DP operation. The main purpose of CGHF is to gather raw information and generate context which is distributed in a standardized way across the different NFs deployed. The CGHF also takes into account the possibility of handling and sharing the context with external nodes, such as applications and third party services. These aspects will be detailed in Section 3.

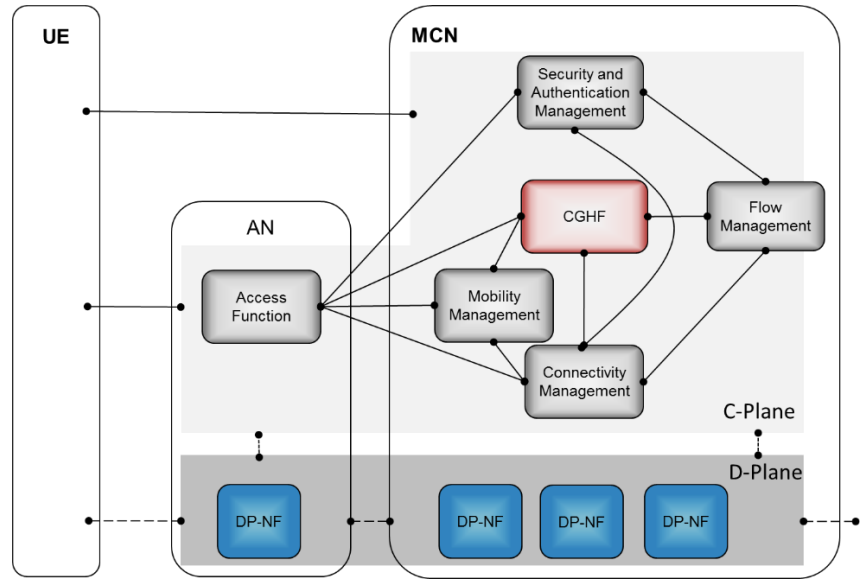

*Figure 2 - Next Generation modularized MCN*

# 3 Context Awareness Framework in MCN

In this section we present the context awareness framework (CAF) the main component of which is the context generation and handling function. In addition we provide insight into the mechanism for exposing context to third parties services via NBI.

## 3.1 Key design Principles

At the heart of any context-aware solution is the principle of *flexibility, i.e., the adaptation* to ever changing situations [11]. Thus, the design will need to identify the operation that is meant to be context aware as well as the contextual information to adjust against. For MCN, such operations could be service anchoring operations, or mobility management, while the range of possible contextual information used for a flexible adjustment of these operations can include network as well as user level information, i.e., what we call raw information. While the first step in the design of a context awareness framework is identifying the operations to adjust, the *support for varying definitions of context* is the next one. If operating in an environment where the concepts for context are well known and/or well defined, one can assume a single context ontology that formalizes these concepts. In such cases, the system ontology will form the basis for the flexibility within the context-aware system.

In addition to ontologies, it is important to consider the facts that provide the basis for deriving context and are obtained by processing raw information. Given the possible wide range of context, it is important to provide suitable means for handling the raw information fed into the CAF, since the source of this information can arise at different timescales and at distributed locations in the communication system. Thus, the gathering of raw information, which will be turned into facts for a CAF, must support means for *spatially and temporarily decoupling*. Finally, the *knowledge creation mechanisms* that often employ self-learning methods, should allow for the system to enrich the context based on the increasing supply of facts. We take these principles into consideration in our solution, as presented next.

### 3.2 CONTEXT GENERATION AND HANDLING FUNCTION ARCHITECTURE

Context Generation and Handling Function (CGHF) is responsible for generating the context as well as managing it. In this work, we propose CGHF as a specific CP function in order to make the context reusable and interoperable, while adhering to the principle of modularization discussed in Section 2. As shown in Figure 3, CGHF essentially gathers the *raw information* from the UE, Radio Access Network (RAN), NFs in the labelled as NF1, …, NFn, and even from third party services and applications via NBI. This raw information is then converted to *facts* by the *Fact Generator* and then ultimately converted to meaningful knowledge (*context*) by the *Context Generator*. The created context is then consumed by the entities which subscribe to it in order to optimize their operations. This includes NFs in CN, applications or even third party services. While the NBI is used for exposing context information, it is also used for receiving information streams which are utilized by CGHF when generating context. It is also important to highlight here the block labelled *peer CGHF* which can be in another slice or of a different provider. So the architecture proposed in this work also takes into account the peering possibility with other CGHFs for exchange of context.

There are two key tradeoffs which need to be considered in the CGHF design. First is the temporal aspect which dictates the time scale at which the CGHF can influence the decision of the involved entities. For instance, handover execution by the mobility management network function must be performed in a very short time scale (in milliseconds) regardless of any other situation of the network (e.g., AN and CN load). Meanwhile the determination of best options for mobility anchoring point selection can be performed over longer periods of time (e.g., in seconds or minutes) according to the situation of the network. The CGHF proposed in this work can handle both these time scales but the focus here is to support the operation of NFs in a longer time scale as the short term decision making can be handled locally by the NFs themselves.

The second aspect is related to the performance on generating context. Having a centralized CGHF with a global view of the system can allow for global optimizations. Nonetheless, there is a tradeoff between such view and generation of timely context. This relates to the validity or expiry of the monitored raw information and facts which constitute the context. There is a high cost on transferring the raw information to the centralized CGHF and analyzing all the facts generated in the network. To this end, we foresee a distributed implementation of CGHF to enable high performance and timely generation of context. In such a scenario the CGHF instances can run in specific regions to locally solve the identification of context and via collaboration amongst each other exchange information to guarantee also the possibility of generating context that are not only regionalized but are covering multiple regions. We next present in detail the architecture of CGHF as illustrated in Figure 3, which is composed of four major components: *Pub-Sub Engine*, *Fact Generator*, *Context Generator*, and *Context Manager*.

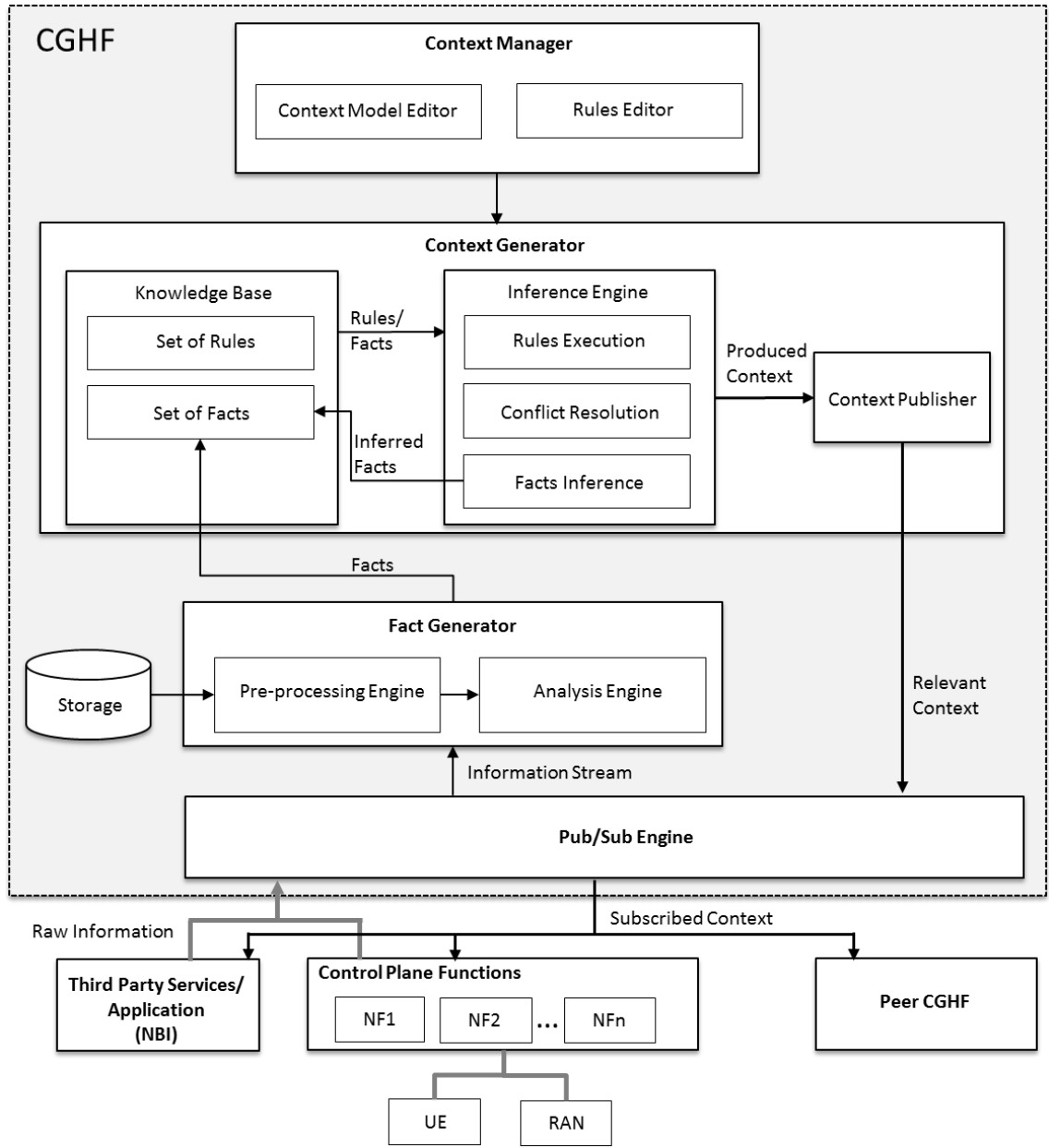

*Figure 3: Logical Architecture of the Context Generation and Handling Function (CGHF)*

**Pub/Sub Engine**

The Publish/Subscribe (Pub/Sub) engine is bidirectional. One the one hand it is responsible for collecting raw information from different sources for context generation while on the other hand it is also used to expose the relevant generated context to the consumers which subscribe to it. The exchange of information in the Pub/Sub engine is realized through a messaging pattern where publishers do not program the messages to be sent directly to specific subscribers, but instead characterize messages into classes without knowledge of the subscribers. This allows information from different entities of 5G MCN architecture to be disseminated and consumed without the need of interfaces among all entities.

**Fact Generator**

The raw information arriving at the Pub/Sub engine is transformed into *Information Streams* that are forwarded to the Fact Generator. The main functionalities of the Fact Generator are supporting data processing and the fact generation. A fact is an assertion about a variable or information. Examples of facts are: #AP1 density is above 90%; #AP1 connected to gateway #Z; average jitter at gateway #Z increase 20% in the last 30 minutes. The Fact Generator, via *Pre-processing Engine*, enables data fusion (e.g. averages, rate of change) so to prepare the information streams for the actual generation of facts. The pre-processed data is then analyzed by a set of algorithms for forecasting, classification, clustering, and trend analysis purposes at the *Analysis Engine*. In addition, the Fact Generator interacts with a *Storage* database, where historical data is kept and made available for pre-processing. This Storage database also supports the generation of facts that are associated with information streams temporally decoupled, such as average of values in the last 30 minutes. The generated facts are then transferred to the Context Generator.

**Context Generator**

The *Context Generator* is responsible for supporting reasoning over the available facts, context models and rules. It consists of (i) the *Knowledge Base* storing the Set of Facts and Rules, and (ii) an *Inference Engine* that supports *Rule Execution* (i.e., reasoning), *Conflict Resolution* over the provided set of facts and rules as well as *Fact Inference*, which are feed back to the *Knowledge Base*. Upon inference, the *Produced Context* is sent to the *Context Publisher* that transforms it into the *Relevant Context*, which is a topic to be pushed into the Pub/Sub Engine. One key difference of the Context Generator proposed in this work to the other reasoning solutions is that the enforcement of changes is not done by Context Generator or CGHF overall. This is because the NFs themselves are in a better position to enforce the changes. Thus, the only action associated with the Context Generator is to publish the relevant context into the Pub-Sub Engine which can then be picked by the entities which subscribe to it.

**Context Manager**

The Context Manager encompasses the Context Model and Rules Editors. The former describes the specific collection of facts and rules that need to be evaluated by the Context Generator in order to identify the occurrence of a relevant context. The Rules Editor are an Event-Condition-Action structure, where the Event-Condition part is associated with the set of facts to be analyzed together, and the Action part describes which relevant context should be published by the Context Generator once the Even-Condition evaluates true.

## 3.3 CONTEXT EXPOSURE VIA NBI

One of the main use of the CGHF will be in customization of the so called network slices, for different vertical services. Hence, there is a need for feature exposure and clearly defined interfaces towards the verticals and applications. To this end, the NBI shall provide services through Application Programmable Interface (API) to share context between stakeholders (e.g., MCN operator and vertical slice operator) and/or orchestrate 5G slices. The main type of API role we foresee for the control plane is to share context. On the one hand, shared context will be used to manage the behavior of 5G network services for control plane requirements during a slice execution with stakeholder objectives (e.g., change the flow policy). On the other hand the shared context will be used to enhance stakeholders services with network information (e.g. network load, user identity) or to optimize the Quality of Experience (QoE) with service level information (e.g. service performance). The API exposure framework for the context exposure NBI with the stakeholders is depicted in Figure 4.

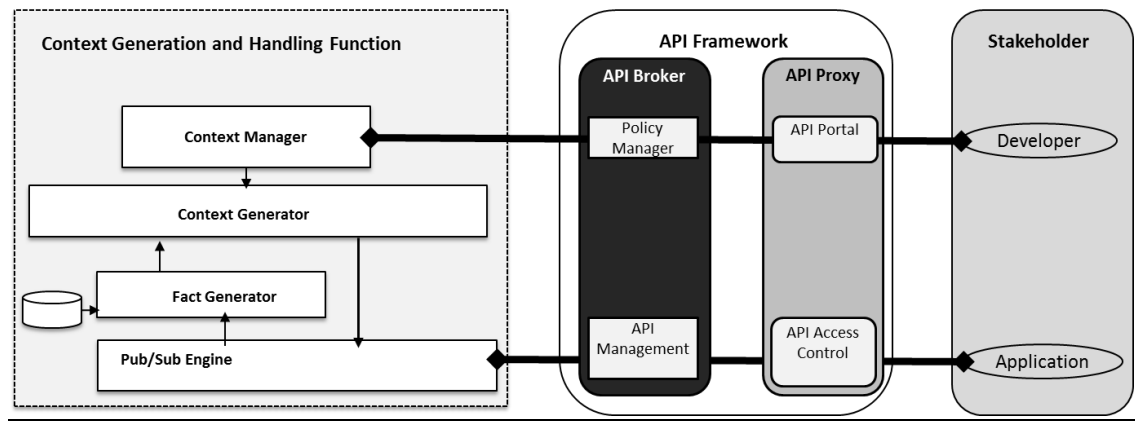

*Figure 4: API exposure framework*

The main components of API exposure framework are: *API Proxy* which is composed by *API portal* (to provide the client interface for the service developer of the stakeholder) and *API access control* (to secure request from third party and stakeholder). The *API broker* consists of *API management* and *Policy Manager in order to provide service tools for API usage by stakeholder.*

The context subscribed by the stakeholder and the exchange of stakeholder information with CGHF are done through the API for NBI exposed by the Pub/Sub Engine. The developers, at design time through the API Portal, will provide all the required information to develop, secure, and invoice both context and information they want associated with their developed services running with the support of the CGHF. After this, the Policy manager of the API Framework configures the CGHF to: process the context required and/or to include in the information exposed by the developer as part of the raw information set that can be used for identification of relevant context. While API Portal and Policy Manager are used at the design time; the API Access control and API Management are used during runtime when either raw information needs to be propagated from stakeholder application to the CGHF or when a relevant context needs to be sent from CGHF to the application. In the next section, we present examples of context utilization, where both NFs and stakeholder application information are used for the detection of relevant context.

## 4 CONTEXT UTILIZATION

Due to rigid and pre-defined architecture of the conventional MCN (Figure 1 (a)), there is not much space for network self-adaptation and optimization. The introduction of softwarization and modularization in 5G opens the possibility of using context awareness to optimize the operation of MCNs. In this session, we present some scenarios of CP operation that can benefit from using relevant context on their mechanisms of decision making. We summarize the relevant contexts used by these scenarios in Table 1.

*Table 1- Relevant Context for Decision Making of CP Procedures. The first columns identifies a relevant context and it is used for reference in the text. The second column describes the relevant context example, which is indeed a description of a situation. The third column lists the type of raw information needed to be generate the facts and infer the relevant context, while the fourth column indicates the source of the raw information. Finally, the fifth column identifies which kind of CP procedures subscribe to receive the relevant context and will enforce changes in the system based on the received context.*

| | **Example of relevant context** | **Raw Information** | **Source of Raw Information** | **Related Control Plane Procedure** |
|---|---|---|---|---|
| 1 | The network detects an unexpected congestion in area Y (e.g., stadium during concert) | Aggregated throughput of Cells in region Y | Access NF | The Policy Function re-evaluates policy decisions based on the context information received and updates policy information of the corresponding network functions: e.g. attributed QoS per subscriber or service is constrained. |
| | | Core network throughput of NFs | NF responsible for monitoring D-Plane Status | |
| | | UE throughput at Cells in region Y | UE | |
| 2 | Application X on device Y is about to suffer more than normal latency | Cell Density | Access NF | Data Plane Anchoring Point Reselection according with Application Type |
| | | Current condition of AN | Access NF | |
| | | Device Location | UE | |
| | | IP anchoring Gateway network status (load, jitter) | NF responsible for monitoring anchor point | |

| | | Actual observed latency in DP Function | NF responsible for monitoring D-plane status | |
|---|---|---|---|---|
| 3 | User experience in service X is lower than required by application | UE resources and processing capabilities | UE | Relocation of Service Point |
| | | User mobility – Device location with respect to cloud server | UE, Access NF | |
| | | Network element (i.e., data center) processing capabilities to accommodate service VM and load information (headroom, jitter, delay…) | NF Responsible for monitoring data centers | |
| | | D-Plane latency | NF responsible for monitoring flow statistics | |
| 4 | The network is about to identify a better network point of attachment from different access technology reported by UE X. | Technology ID and key link characteristics (i.e., bandwidth, channels, QoS classes) | Access NF | Trigger UE traffic redirection or handover towards new access technology Point of Attachment |
| | | Supported Protocol List (i.e., network layer protocol, mobility support, P2P-enabled) | Access NF, UE | |
| | | Neighbor access point link conditions (i.e., load, maintenance, deployment conditions) | Access NF | |

## 4.1 CONTEXT UTILIZATION FOR POLICY CONTROL

The Policy and Charging Control (PCC) architecture as standardized in 3GPP [5], defines a framework that allows operators to control and tailor data network access setup based on: individual (or group) subscription profiles; external (third-party) specific requests; and network status occurrences, either specific to individual data connections or a certain number of data connections, e.g., sharing a specific link or cellular area. To enable the latter functionality, the PCC policy decision point, the Policy and Charging Rules Function (PCRF), is able to receive network originated information, like a user location change, different access type coverage, or the status of associated consumption (traffic volume or time based). In pre-5G 3GPP releases, network-status information can also be obtained from the RAN Congestion Awareness Function (RCAF), which provides the PCRF with congestion status information on a certain area for all or specific subscribers. The PCRF, thus, can force the PGW, for example, to change the anchor point or downgrade QoS characteristics. More dynamic and programmable adaptation of the PCRF has also been studied [12]. However, in all these cases, network context is gathered for a centralized PCRF decision making on data connectivity setup, and then enforced in a centralized node as the PGW.

In 5G however, the fundamental functions instantiated by the gateway are separated into CP and DP virtualized functions, perhaps with multiple instances as illustrated in Figure 1 (b), which either may not be visible to the PCRF or would require an extra design to consolidate all the information from the multiple instances. Therefore, we propose the move from the information collection as is made nowadays by the PCRF towards a model where a Policy Function [8] (the 5G version of the PCRF) receives relevant context from the CGHF. For instance, the first example in Table 1 depicts a relevant context definition associated with such Policy Function for the detection of congestion situation. Once the Policy Functions receives such context, it can update all the policy-related information to constrain or adapt the QoS characteristics, without having to handle by itself the monitoring and detection of the situation.

## 4.2 D-PLANE ANCHOR POINT RESELECTION

The 4G/LTE provides limited mechanisms for selection and reselection of data plane anchoring points, such as Serving Gateway (SGW) and PGW, executed by the Mobility Management Entity (MME) [5]. The selection is performed based on the weight factor associated with options of anchoring point received from the Domain Name Service (DNS) specialized for 4G/LTE, and the load information retrieved from the anchoring points via GPRS Tunneling Protocol for Control (GTP-C) messages. The reselection case can happen due to handover or overloading of the gatways. The shortcomings of current system and information organization in gateway (re)selection are three folds. First, there is no mechanism in PCRF, to monitor the delay budget of a data transmission. The implications are the need for capacity planning and over provisioning of transport networks, since overloaded EPC entities results in degradation of QoE for the users. The lack of DP resources will be identified by the EPC management system and not by the EPC control entities, which operates in different temporal cycles thus delaying any reaction to solve bad QoE. Second, selection of gateways taking into account the UE type currently requires dedicated pool of gateways only to serve that type of UEs. This prevents an efficient load balancing among all available gateways in a given region where different UE types need to be served, leading to inefficient use of network resources, thus reflected in the CAPEX/OPEX. Third, the GTP-C overload control uses control signaling information to indicate that a gateway is overload. When the MME uses the GTP-C information, it removes simultaneously the CP and DP of overloaded gateways, these planes of gateway in EPC are coupled in a physical device. In 5G MCN, CP and DP are decoupled, thus without explicit information from DP gateway only CP signaling will not guarantee that a gateway can actually handle the load directed to it. In addition, the DP gateway function will not be implemented in hardware but in software. The variation of performance both in terms of processing load as well as traffic amount cannot be ignored.

Therefore, in 5G, triggers and information used by CP functions for reselection of anchoring points need to evolve as the system evolves, and using context. For instance, the second example in Table 1 shows the relevant context that can trigger the gateway reselection based on enriched context. This context combines information, about AN load, UE application performance, and status of the anchoring points of the 5G network. By identifying the applications of UEs with QoE degradation, the CP NF can take decisions to optimize the assignment of gateways so that the QoE of the UE's applications improve, as well as guaranteeing a better utilization of the network resources.

### 4.3   SERVICE POINT RELOCATION

A *Service Point* is a cloud-computing platform placed in a network element that enables or assists a service provided by a third party. The functionalities running on service points range from data analytics, computing platforms for different vertical services, or execution point for services such as a video codec [13]. 3GPP standards have not included service points in their architecture, and DP services are always held outside of PGW [8], as illustrated in Figure 1 (a). There have been attempts to provide access to localized content with the introduction of the concept of Local Network and Selective IP Traffic Offload (SIPTO) function in 3GPP [5]. Nevertheless, connection to a Local Network does not necessarily guarantee access to specific content unless the Service Point is proprietary collocated within the local network.

There are multiple motivations and benefits of including service points within the MCN network, as illustrated in Figure 1 (b), like enabling lower end-to-end latencies and ultra-responsive networks. Context awareness can be foreseen as an enabler for optimal placement of service points. Even if at placement time a service point is deemed to be in an optimal position in relationship to its users, the mobile network conditions change over time. The third example in Table 1 shows a relevant context to indicate a situation in the network which requires service point relocation. By combining the information about the UE's application and mobility pattern, with the load of the entities related to the service point as well as the DP load of MCN, it is possible to recognize opportunities to optimize the mapping among UEs and service points, therefore improving the perceived QoE. In this sense, service relocation triggered by the context information can influence the CP functions related to the user connectivity and more precisely, the end-to-end path management.

## 4.4 MULTI-ACCESS CONTEXT

Connectivity optimization is an essential aspect of MCN CP procedures, becoming even more important with current scenarios where the same UE has multiple access technologies. However, the fundamental specificities of the way different access technologies are used and controlled, creates an opportunity for optimizing UEs connectivity that, despite being recognized for over a decade, has not been harnessed to its full potential [14]. There is a lack of seamless traffic or attachment transition to an alternative point of the network in an optimized way. As a result, solutions from different standardization bodies are not aligned and focus on specific technologies, while following non-convergent strategies (i.e., technology integration). For example, 3GPP focuses on offloading techniques towards Wi-Fi, and IEEE (with 802.21 and 802.1cf) focuses mostly on inter-IEEE technology interaction. As a result, this disparity prevents the converged utilization of contextual link information from different access technologies [15], which limits the potential to optimize connectivity procedures. In Table 1, the fourth example shows how a relevant context can be used to optimize the use of different access network technologies, by changing the access attachment point of an UE to use the best access for the UE. In this case, the identification of such relevant context uses UE are parameters of interfaces, detected network connection points, current link conditions, the RAN available capacity, connected nodes, and supported features. The CGHF assumes the role of providing the right level of awareness of access-specific information to CP decision entities, such as Access Function (illustrated in Figure 2).

## 5 CONCLUDING REMARKS AND CHALLENGES AHEAD

Endowing 5G control plane with context awareness enables the refinement and definition of new procedures that can both optimize the execution of the NFs at the control plane and improve the QoE of users. To this end, we present a network architecture with a logically centralized entity, Context Generation and Handling Function (CGHF), as to collect various information, i.e. raw information, from the network, and analyze those so to provide knowledge, i.e., relevant context that can be used by different optimization processes. We elaborate some examples of such, with details of the relevant context and the potential benefits, in this paper.

Nonetheless, there are number of challenges that need to be addressed, including, but not limited to the followings: (1) How to benefit from context reusability while handling the occurrence of conflicting changes in the system triggered by different NFs or applications based on the same context; (2) How to efficiently expose the monitored information from multiple sources to the logically centralized CGHF is yet another challenge; (3) It is also important to design the monitoring and collection of information at the CGHF so that it does not overwhelms the operation of the MCN; (4) The alignment of the temporal scales of the monitored information used to infer relevant context and the "freshness" of such inferred context, is another important aspect mainly because different NFs and applications operate in different time scales.

**Acknowledgement**

The authors want to Marco Liebsch, Panagiotis Gouvas, Qing Wei, Benoit Radier, Ulises Olvera-Hernandez, Riccardo Trivisonno, and Frédéric KLAMM, which are members/partners of the project as well researchers and standardization delegates in- and outside the project for the very fruitful discussions. The cooperation COntrol Networks in FIve G (CONFIG - http://www.5g-control-plane.eu/) is not funded by the international and national funding authorities and runs on a self-funding basis.